\documentclass{aa}  

\usepackage{graphicx}
\usepackage{color}
\usepackage{txfonts}
\newcommand{\kms}{\mbox{km\,s$^{-1}$}}
\newcommand{\new}[1]{{#1}}

\begin{document}

   \title{Non-thermal line broadening due to braiding-induced turbulence in solar coronal loops}



   \author{D.~I.~Pontin
          \inst{1,2}
          \and
          H.~Peter\inst{3}
          \and
          L.~P.~Chitta\inst{3}
          }

   \institute{School of Mathematical and Physical Sciences, University of Newcastle, Callaghan, NSW 2308, Australia\\
              \email{david.pontin@newcastle.edu.au}
           \and
             School of Science and Engineering, University of Dundee, Nethergate, Dundee, DD1 4HN, UK\\
                  \and
             Max Planck Institute for Solar System Research, Justus-von-Liebig-Weg 3, 37077, G{\" o}ttingen, Germany\\
     }

   \date{\, ; \,}

 
  \abstract
   {}
   {Emission line profiles from solar coronal loops
exhibit properties that are unexplained by current models. We investigate the non-thermal broadening associated with plasma heating in coronal loops that is induced by magnetic field line braiding.}
   {We describe the coronal loop by a 3D magnetohydrodynamic model of the turbulent decay of an initially-braided magnetic field. From this
we synthesize the \ion{Fe}{xii} line at 193\,{\AA} that forms around 1.5\,MK.}
   {Key features of current observations {of extreme UV lines from the corona} are reproduced in the synthesised spectra:
(i) Typical non-thermal widths range from 15 to 20 \kms. 
(ii) The widths are approximately independent of the size of the field of view.
(iii) There is a correlation between line intensity and the non-thermal broadening. (iv) Spectra are found to be non-Gaussian, with enhanced power in the wings of order 10-20\%. }
   {Our model provides an explanation that self-consistently connects the heating process to the observed non-thermal line broadening.
The non-Gaussian nature of the spectra is a consequence of the non-Gaussian nature of the underlying velocity fluctuations,
interpreted as a signature of intermittency in the turbulence.}


   \maketitle
%

\section{Introduction} \label{sec:intro}
\subsection{Observational basis}

Observations of spectral lines on the Sun almost always reveal the presence of non-thermal broadening of the line profiles, irrespective if it is an absorption line in the near-surface layers of the photosphere and chromosphere \cite[e.g.][]{2017SSRv..210..109D} or an emission line from the transition region and corona {\cite[e.g.][and references therein]{1992str..book.....M,delzanna2018}}.
%
%
{Specifically,} the observed width of an emission line, $w_{\rm{obs}}$ (corrected for instrumental broadening), {is generically found to be in excess of the thermal width, $w_{\rm{th}}$.
The thermal width represents the thermal motions of the plasma in the source region of the emission.
In equilibrium, this equals the width of the velocity distribution (Maxwellian) of a gas at temperature $T$ \cite[e.g.][]{1992str..book.....M},}
\begin{equation}\label{E:th.width}
w_{\rm{th}}=\left(\frac{2\,k_{\rm{B}}T}{m}\right)^{\!\!1/2},
\end{equation}
Here $k_{\rm{B}}$ is Boltzmann's constant and $m$ is the mass of the atom, ion or molecule under consideration.
{When quoting the {thermal} width of a given spectral line the temperature usually used is the line formation temperature {assuming ionisation equilibrium.}}
Here and in the rest of this study we refer to the Gaussian width which is the half width at $1/e$ of the line peak.

Due to the Doppler effect, the observed line profile should have this same thermal width $w_{\rm{th}}$.
The simple fact that the observed line widths are in excess of the thermal broadening, $w_{\rm{obs}}>w_{\rm{th}}$, shows that there are non-thermal motions that are not resolved by the observations.
Assuming that these non-thermal motions also follow (roughly) a Maxwellian distribution, one can attribute a non-thermal width, $w_{\rm{nt}}$, to these,
\begin{equation}\label{E:non-thermal}
w_{\rm{nt}}=\big(w_{\rm{obs}}^2-w_{\rm{th}}^2\big)^{1/2}.
\end{equation}
The non-thermal motions can originate from {any} unresolved motions within the spatial resolution element of the instrument:
{One example would be that there is a structure on the Sun at scales below the resolution limit of the instrument, so that plasma at different locations within the resolution element would move at different (line-of-sight) velocities.
Also, there might be different speeds at different locations along the line of sight.
Or the velocity could change in time during the exposure of the spectrum.
On the real Sun we are probably confronted with a mixture of all these effects.
The non-thermal broadening typically corresponds to unresolved motions at subsonic speeds \cite[e.g.][their Fig.\,5.2]{1992str..book.....M}.
As such the motions could in principle be associated with {turbulence \cite[e.g.][]{doschek1977},  quasi-periodic upflows or waves \cite[e.g.][]{tian2012} or} shocks \cite[e.g.][]{2015ApJ...799L..12D}
%
%
or by any subsonic motions that are unresolved.
}
Many speculations have been made regarding the physical nature of these non-thermal motions, but so far it has not been possible to draw any firm conclusions.
Instead, the use of, e.g., terms such as micro- or macro-turbulence implemented in atmospheric modeling \cite[][]{1978SoPh...59..193G} reflect more the current lack of understanding than a clear physical understanding of the process.

%

The extreme ultraviolet (EUV) radiation from the Sun is dominated by emission lines originating from the upper solar atmosphere, from the transition region and the corona hosting plasma at temperatures from 0.05\,MK to several MK.
Since the early EUV observations it has been obvious that all these lines show significant non-thermal broadening {\cite[][]{1972spre.conf.1595B,1973A&A....22..161B,doschek1977}}.

In quiet Sun regions the non-thermal line width increases with temperature up to about 0.3\,MK reaching some 30\,{\kms; the broadening then drops to about 15\,{\kms} just above 1\,MK} \cite[][]{1992str..book.....M,1998ApJ...505..957C}.
Modest values of non-thermal broadening of about 15\,{\kms} to 20 \,{\kms} are also found in active regions in loops at high temperatures ranging from {1\,MK to 5\,MK} \cite[][]{1999ApJ...513..969H,2016ApJ...827...99T,2016ApJ...820...63B}.
{Magnetically open structures in coronal holes might show slightly higher broadening \cite[e.g.][]{2013ApJ...763..106H}, but there the broadening mechanism might be different from the magnetically closed loops we consider in this paper.}
One remarkable feature of the non-thermal broadening is that it does not change with spatial resolution, from the smallest scales currently observable  (about 0.35{\arcsec}) to several arcseconds.
This is true for transition region temperatures of 0.1\,MK \cite[{seen in \ion{Si}{iv}};][]{2015ApJ...799L..12D} as well as for coronal structures above 1\,MK  \cite[{seen in \ion{Fe}{xii}};][]{2016ApJ...827...99T}. 
This shows that the mechanism leading to the excess broadening has to operate on spatial scales smaller than about 0.35\arcsec.

{In the transition region at temperatures of about 0.1\,MK, the non-thermal broadening increases or correlates with the line intensity \cite[][]{1984ApJ...281..870D}, i.e. the non-thermal broadening is larger for higher line intensity.}
This is the case~irrespective of whether we consider quiet Sun, coronal holes, active regions \cite[][]{2015ApJ...799L..12D}, rapidly evolving microflares in active region cores \cite[][]{2020ApJ...890L...2C}, or the bright network or faint inter-network areas \cite[][]{2000A&A...360..761P}.
At higher temperatures in the quiet Sun the correlation becomes  weak, if present at all \cite[][]{1998ApJ...505..957C}.
In active regions at temperatures above 1\,MK there is also a correlation between line intensity and non-thermal broadening if the active region is considered as a whole \cite[][]{2009RAA.....9..829L}.
Also when considering a single coronal loop there is a (weak) correlation  \cite[{seen in \ion{Fe}{xii}};][]{2016ApJ...827...99T}.
In a few selected locations in an active region, e.g.\ when picking a small particular segment of a loop or a footpoint region, one might even find a negative correlation \cite[][]{2011ApJ...742..101S}.
So in general, in an active region the non-thermal broadening should increase with line intensity \cite[except for the dark outflow regions in the periphery of active regions][]{2008ApJ...686.1362D}.

Besides the non-thermal broadening, the spectral profiles often also show excess emission in the line wings.
This was already noted in early EUV observations \cite[][]{1977ApJ...211..579K}.
The more prominent of these cases that often show separate components in the wings have been interpreted as bi-directional outflows from reconnection jets \cite[][]{1993SoPh..144..217D,1997Natur.386..811I,2015ApJ...813...86I}.
In particular, lines forming in the quiet Sun transition region at around 0.1\,MK show an excess emission in the line wings contributing in some cases up to 30\%  to the total line radiance \cite[][]{2001A&A...374.1108P}, often with an excess to the blue \cite[][]{2009ApJ...707..524M}.
A clear excess emission in the blue wing is also found at the footpoints of coronal loops in active regions \cite[][]{2008ApJ...678L..67H}, which has been interpreted as the signature of a universal process that injects mass and energy into the loop \cite[][]{2009ApJ...701L...1D}.
If stable coronal loops are observed near the apex with a line-of-sight  apparently (roughly) perpendicular to the loop axis, the excess in the blue and red wings is symmetric \cite[{seen in \ion{Fe}{xv}};][]{2010A&A...521A..51P}, i.e.\ of comparable strength.
This is the case not only for these active region loops at more than 1\,MK, but is also found for cool active region loops at about 0.1\,MK \cite[{seen in \ion{Si}{iv}};][]{2019A&A...626A..98L}.

\subsection{Previous hypotheses and modelling}
Many {qualitative hypotheses have been proposed}  to explain the non-thermal line widths, but so far we have no clear quantitative explanation \cite[e.g. as summarised in Sects 4.3--4.5 by][]{2010A&A...521A..51P}.
{Such qualitative} suggestions to explain the line broadening and the excess emission in the wings include magneto-acoustic waves and shocks, field-aligned non-resolved motions, small-scale reconnection, or turbulent motions or kinetic processes leading to modifications of the velocity distribution function.

{More quantitative predictions of the line profiles have recently been developed {based on waves injected into the loop} -- however, they fail to simultaneously reproduce both observed temperatures and non-thermal broadening.}
{On the one hand, 3D MHD models that excite transverse MHD waves through velocity drivers with prescribed amplitudes find that the line-of-sight superposition of large-amplitude Alfv\'enic waves in the coronal loops could result in a large non-thermal broadening of the spectral lines \cite[][]{2019ApJ...881...95P}. However, those models {do not simultaneously heat the plasma with those waves {to the required temperatures}. Instead, a rather hot {background} already exists in the initial condition.}
On the other hand, reduced MHD models that do self-consistently energize the coronal loops through Alfv\'enic waves} {do not match the observed temperatures and non-thermal broadening simultaneously: 
\new{if the non-thermal broadening is matched to observations then temperatures are too low, while in simulations that reproduce the observed temperatures the non-thermal broadening is too high}
\cite[][]{2014ApJ...786...28A,2017ApJ...849...46V}.}
As noted above, typical observed non-thermal broadening is in the range 15--20 km s$^{-1}$, while in these models \new{typically $w_{\rm nth}\sim 27$ km s$^{-1}$.}
{Because of the principle of reduced MHD models, one cannot derive the actual emission line spectra from those simulations \cite[][]{2014ApJ...786...28A}.}

{Also one-dimensional loop models have been used to investigate non-thermal broadening. By design, such models can only investigate the effects of field-aligned flows on the line width. For example, \cite{2006ApJ...647.1452P} investigated the effect of a short heating pulse. The resulting flows can lead to line broadening; in the right range of model parameters even giving the right amount of non-thermal broadening when integrating over the whole loop. However, by construction, such models would always predict zero non-thermal broadening at the apex of the loop if viewed roughly perpendicular to the loop axis. However, this is not observed \cite[e.g.][]{2010A&A...521A..51P}, and an additional mechanism would be required to explain the non-thermal motions perpendicular to the magnetic field.}

In 3D MHD models of an active region the non-thermal broadening does not change significantly with line formation temperature (contrary to observations) and falls short by a factor of two at the temperatures with the strongest broadening of about 30\,{\kms} \cite[][]{2006ApJ...638.1086P}.
{This indicates that the spatial resolution of such 3D active region models is not sufficient: if the resolution were higher, the simulations would produce motions on scales smaller than those resolved in current models, and presumably the velocities on the small scales could be higher. In turn, this could provide an explanation for the observed non-thermal broadening.}
%
To understand the observation that the non-thermal broadening does not change with spatial resolution, \cite{2015ApJ...799L..12D} suggested that acoustic shocks along the line of sight could produce the non-thermal broadening.
While their results gave the correct general correlation between intensity and width, the non-thermal broadening fell short by a factor of two to three.
Also, none of the forward models (i.e.\ those that synthesize emission line profiles) gave systematically enhanced wings of the line profiles as found in observations. 

\subsection{Summary and purpose}
In summary, a number of observed properties of coronal loop emission spectra (forming around one to a few MK) exist to constrain models, and so far remain (collectively) unexplained: (i) the non-thermal broadening is of the order of 15--30 \kms, (ii) the broadening is approximately independent of instrument resolution (down to $\sim$0.35\arcsec), (iii) the correlation of line intensity and the non-thermal broadening, (iv) the line profiles are non-Gaussian with enhanced power in the wings.
{The properties (i), (ii), and (iv) are all common to the emission lines originating from the transition region and corona, e.g. \ion{C}{iv}, \ion{Si}{iv}, \ion{Mg}{x}, \ion{Fe}{xii}, or \ion{Fe}{xv}, and have been seen essentially by all solar extreme UV spectrometers with sufficient spectral resolving power. Only property (ii) has been reported more recently, so it is documentded only for fewer lines (\ion{Si}{iv} at 1394\,\AA; \ion{Fe}{xii} at 195\,\AA\ and 1349\,\AA) observed with two instrument \cite[IRIS, EIS;][]{2015ApJ...799L..12D,2016ApJ...827...99T}.
For a first-step model to provide a self-consistent explanation of these observed properties, we will focus on just one spectral line originating from the corona.
}

In our study we will concentrate on the non-thermal motions as observed in the extreme ultraviolet (EUV) Fe\,{\sc xii} emission line from the solar corona originating at temperatures of about 1.5\,MK.
We investigate synthesised Fe\,{\sc xii} emission line profiles based on simulations of MHD turbulence in coronal loops -- examining the extent to which the above four properties are reproduced. 
The turbulence is induced naturally during the relaxation of a braided magnetic field within the loop, as in the braiding model for coronal heating introduced by \cite{parker1979,parker1988}.
In the following section we describe the simulations and methods of our analysis. 
Then in Section \ref{sec:line} we analyse the spectra obtained, and compare this with the distribution of velocities within the loop. In Sections \ref{sec:discussion} and \ref{sec:conc} we present a discussion and conclusions.
%
%
%


\section{Numerical simulations {and spectral synthesis}} \label{sec:simulations}
We analyse the results of resistive MHD simulations that include thermal conduction and optically thin radiation in the energy equation. The simulation setup and equations solved are described in detail in \cite{pontin2017} -- the initial magnetic field we use is the ``Braid1" field from that paper \citep[see also][]{wilmotsmith2010,pontin2011a}. 
Indeed, our simulation setup is identical, except we use here an increased spatial resolution of $640^2\times512$.

In short, we begin {at $t=0$} with a braided magnetic field that is close to -- but crucially not in -- force-free equilibrium. {The subsequent evolution of this field leads to a state of decaying turbulence, described below.}
{At $t=0$ the plasma has a} uniform density ($2\times10^{-11}$ kg m$^{-3}$) and temperature ($5\times 10^5$ K). Hyper-resistivity and -viscosity are employed in order to minimise dissipation of structures that have sizes substantially larger than the grid spacing.
Magnetic field lines in the domain all connect between the planes $z=\pm 24$ (Mm). These boundararies represent the base of the corona, and here the field is line-tied with the plasma velocity being fixed to zero. The $x$ and $y$ boundaries are periodic.  The field line tangling is confined within a radius of $\sim2$ Mm from the $z$-axis compared to a loop length of 48 Mm, and the axial field strength in the loop is 100 G.

\begin{figure}
\centering
(a)\includegraphics[width=0.26\textwidth]{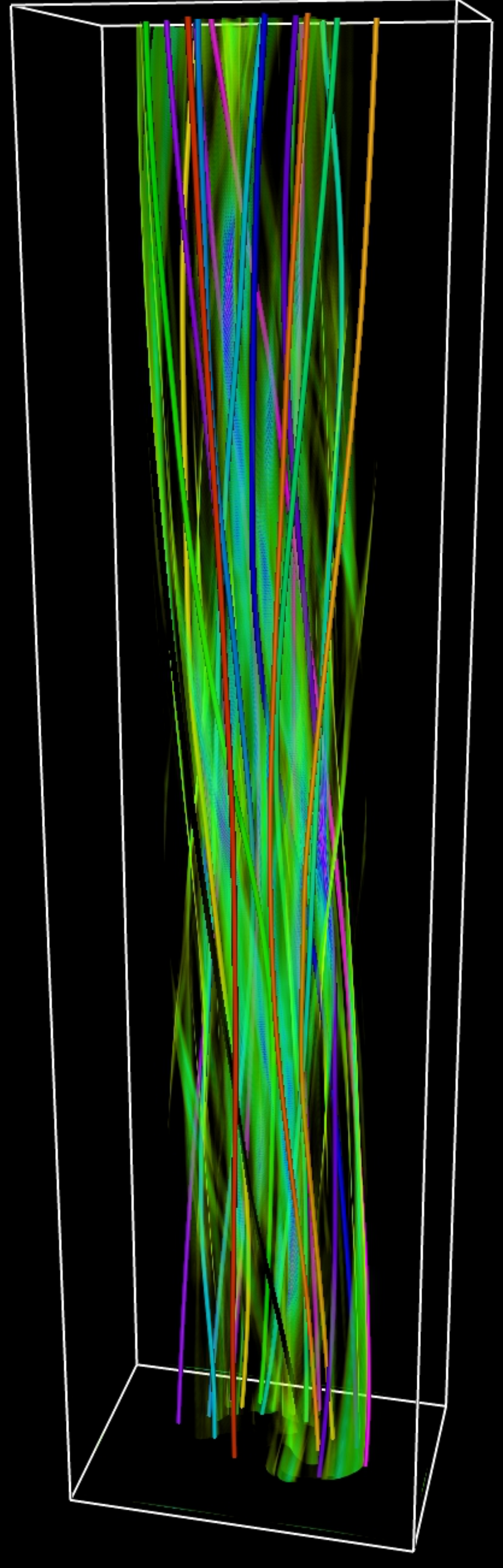}
(b)\includegraphics[width=0.159\textwidth]{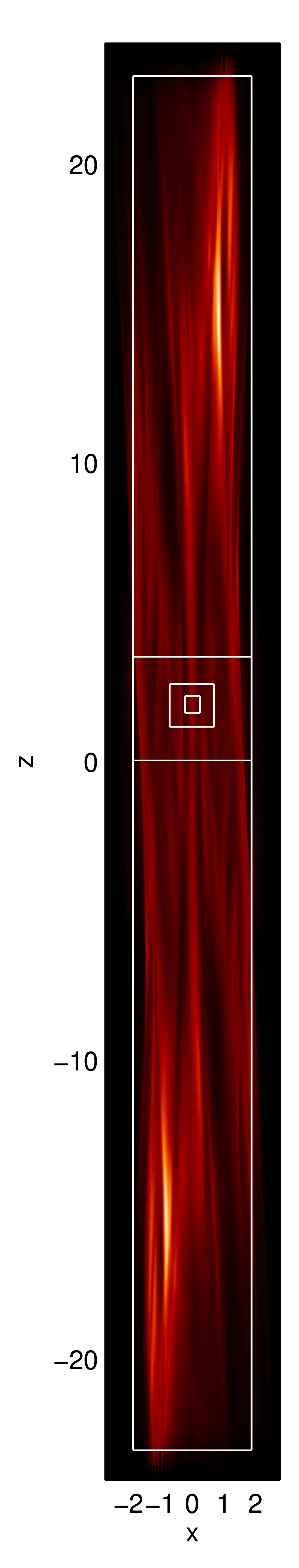}
\caption{Overview of the modelled coronal loop. (a) Volume rendering of the modulus of the current density together with selected magnetic field lines at $t=92$ s. (b) Synthesised \ion{Fe}{xii} emission for the whole loop with LOS along y. Over-plotted are boxes showing the fields-of-view used in Figure \ref{fig:lineprofiles}. Axis markers are in Mm and the intensity is normalised to 1.}
\label{fig:braid}
\end{figure}

The motivation to begin with an already-braided magnetic field -- as opposed to injecting the braiding via boundary motions -- is as follows. It is well established that in flux braiding simulations, the current sheets within the domain grow thinner at an exponential rate as successive boundary perturbations are applied \citep{vanballegooijen1988a,vanballegooijen1988b,mikic1989,pontin2015a}. Therefore the onset of reconnection and energy dissipation is expected to be later for higher magnetic Reynolds number ($R_m$). In particular, for the numerically-accessible values of $R_m$ in 3D MHD simulations\footnote{{In general the value of $R_m$ in simulations is limited by the numerical resolution. Depending on the scheme employed, either an explicit resistivity must be used to stop structures collapsing to the grid scale (and causing numerical artefacts), or numerical dissipation occurs which prevents this collapse to the grid scale, acting like a resistivity. In either case this (effective) resistivity is many orders of magnitude larger than the expected coronal value, meaning that $R_m$ is many orders of magnitude smaller than in the corona.}}, reconnection commences much earlier than it would on the Sun. 
The statistically-steady state that is eventually reached in boundary-driven braiding simulations has an  average field perpendicular to the loop axis, or equivalently degree of tangling, that increases with $R_m$ -- i.e.~again, the field lines are not as braided/tangled at numerically-accessible values of $R_m$ as would be expected at coronal parameters. We therefore take a different approach, which is to begin with a field that is approximately as strongly braided as expected in the corona, based on the estimations of \cite{pontin2015a}, and follow a relaxation process in the presence of dissipation.

\begin{figure}
\centering
\includegraphics[width=0.47\textwidth]{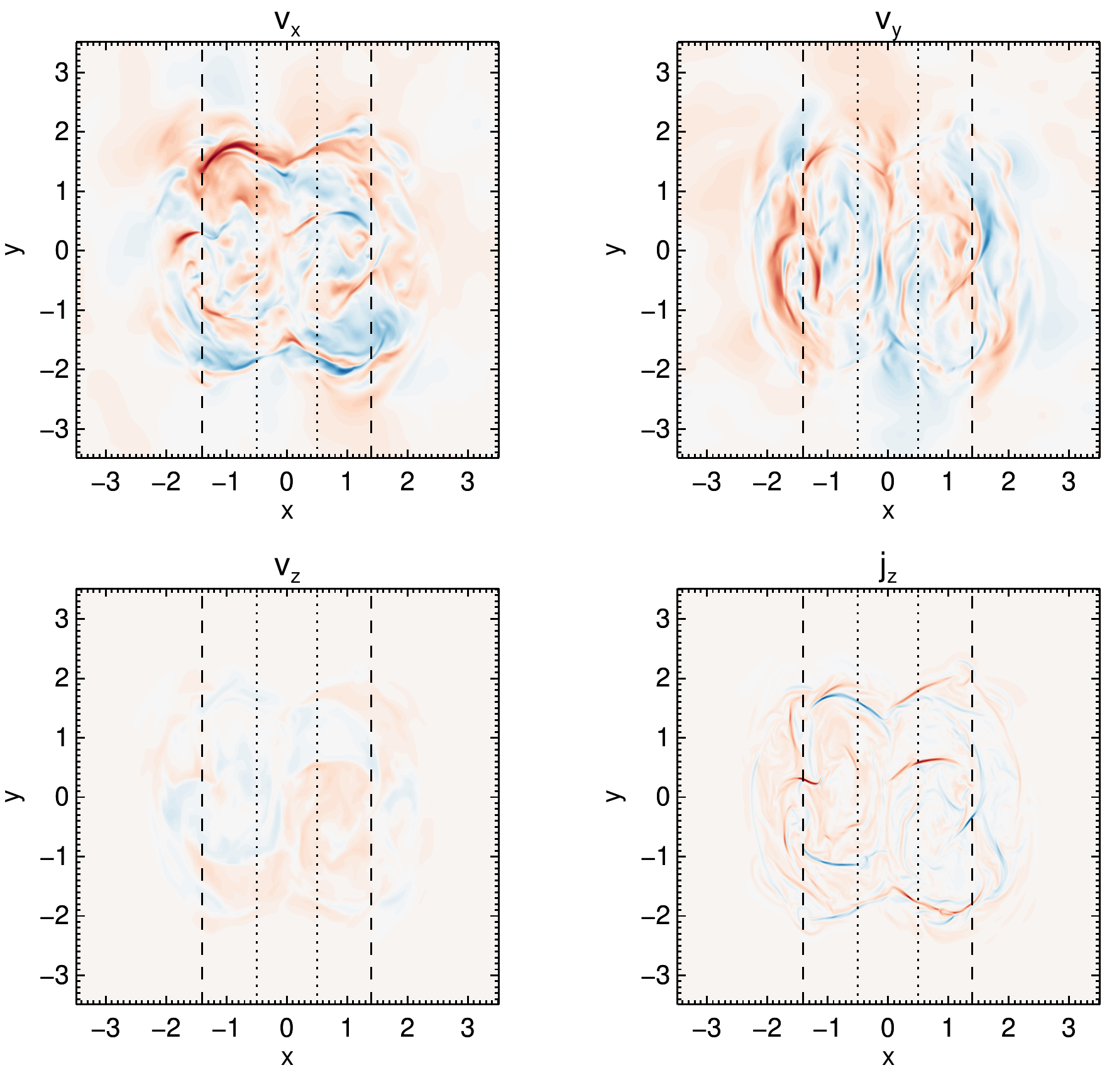}\\
\includegraphics[width=0.23\textwidth]{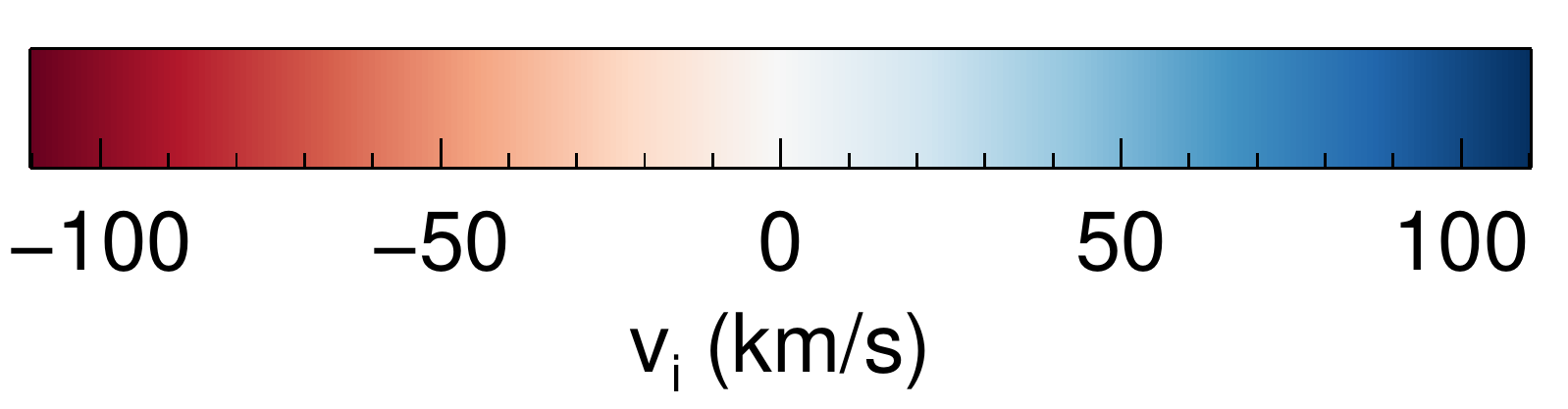}
\includegraphics[width=0.23\textwidth]{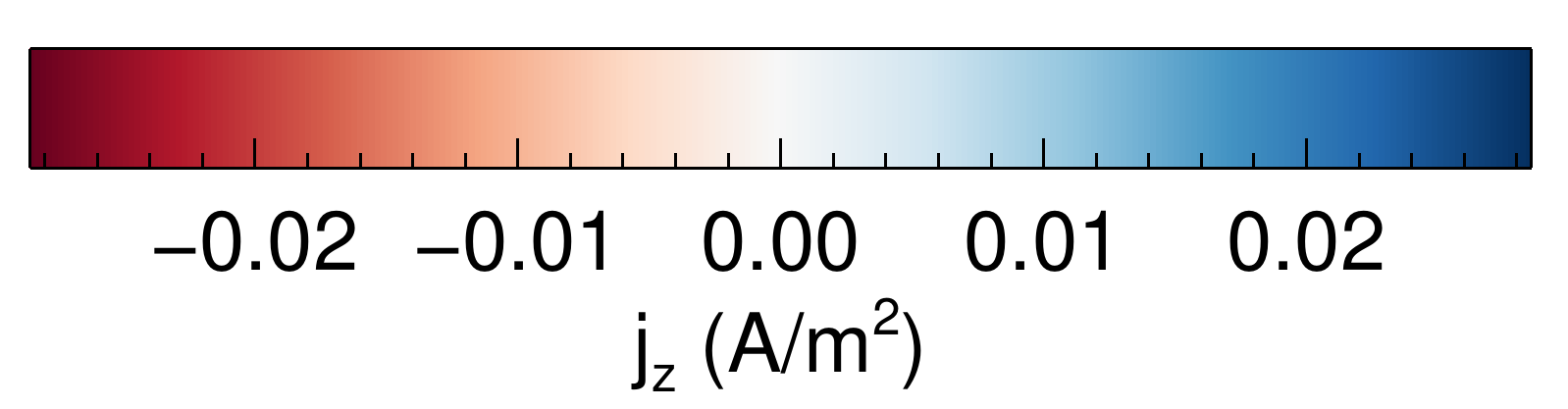}
\caption{Turbulence induced flows and currents in the loop. All three velocity components and the current density parallel to the loop axis, in the plane $z=2$ Mm, after 92 s of the simulation. Over-plotted are boxes showing the fields-of-view used in Figure \ref{fig:lineprofiles}.}
\label{fig:jvcuts}
\end{figure}

The plasma and magnetic field evolution in the simulations has been described in detail in a series of papers \citep{pontin2011a,pontin2016a,pontin2017}, so we give only a summary here.  Due to the highly tangled nature of the field lines, the field rapidly develops thin, intense current layers \citep{pontin2015a}. Reconnection in these current layers triggers a cascade to small scales, fragmentation of the current distribution (e.g.~Figure \ref{fig:braid}), and the initiation of decaying turbulence. During this turbulent decay the magnetic field topology gradually simplifies through many localised reconnection events, with the magnetic energy being converted into kinetic and thermal energy, leading eventually to a non-linear force-free field that retains finite, large-scale twist but no field line tangling \citep{pontin2011a}. Important for the present study are the velocities generated during the turbulent relaxation, which are visualised at a {representative} time in Figure \ref{fig:jvcuts}. As we see from the plots, the velocity fluctuations in $x$ and $y$ are typically larger than those parallel to the loop axis ($z$). This is due to the presence of the dominant $z$-component of the magnetic field.
As discussed in detail in \cite{pontin2011a,pontin2016a}, the turbulence eventually dies out as the thin current layers are replaced by {a weak, distributed current density whose gradient has length scale on the order of the loop width}. At this time the decay of the total energy transitions from power law to exponential in time, and the power-law inertial range in the spatial magnetic energy spectrum is lost. While the system is not in an exact equilibrium the turbulent relaxation is complete -- this transition occurs in our simulations for $t\sim 200-300$\,s.

\begin{figure}
\centering
(a)\includegraphics[width=0.45\textwidth]{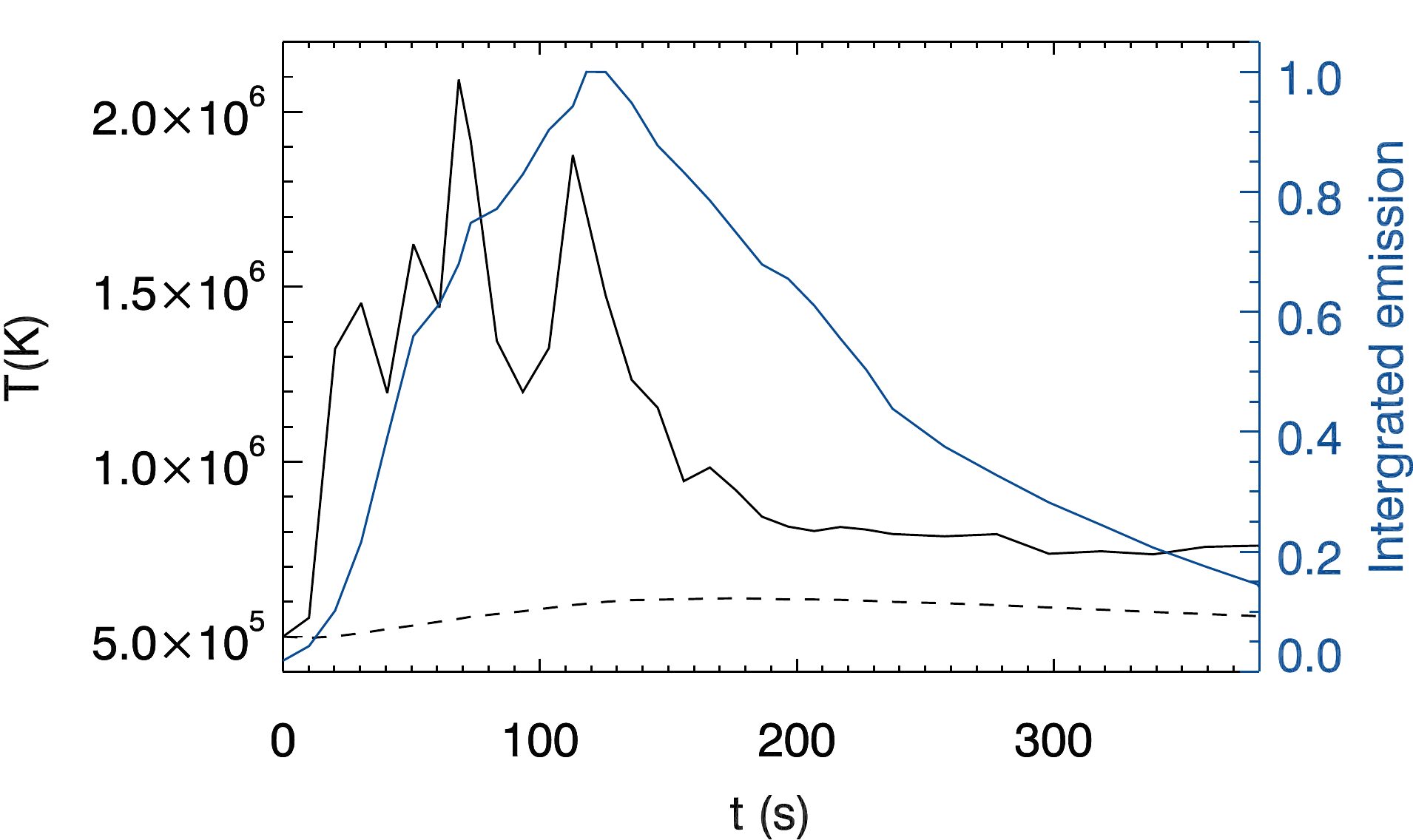}\\
(b)\includegraphics[width=0.45\textwidth]{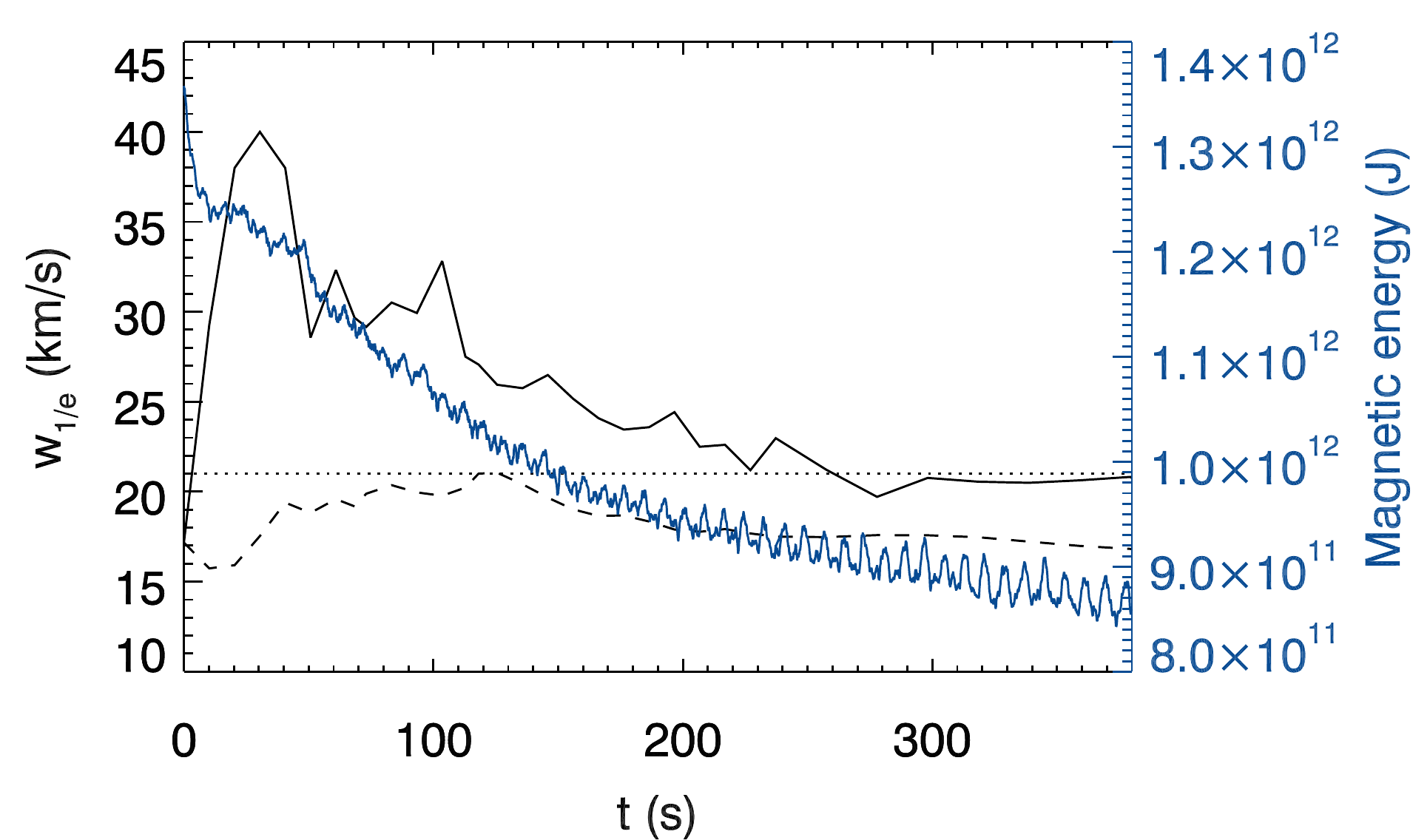}
\caption{Temporal evolution of the relaxing loop. (a) Peak temperature (solid line) and average temperature (dashed) for $x,y\in[-1.9,1.9]$ Mm, $z\in[-23,23]$ Mm, together with the emission in \ion{Fe}{xii} integrated over the same volume, i.e.\ essentially from the whole loop.
%
%
(b) Blue: magnetic energy in excess of the energy of the potential (uniform, vertical) field during the simulation. Black solid line: Spectral line width with line of sight parallel to the y-axis and viewing window with limits $x\in[-1.9,1.9]$ Mm, $z\in[-23,23]$ Mm. Black dashed line: line width with LOS parallel to the loop axis for $z\in[10,20]$ Mm and viewing window $x\in[-2.3,2.3]$, $y\in[-1,2.6]$ Mm. Dotted: thermal width of  21 km/s at  $T=1.5\times10^6$ K.}
\label{fig:lw_em}
\end{figure}

Synthetic emission spectra are calculated using the FoMo package \citep{vandoorsselaere2016}. Line-integrated emission signatures are calculated for the \ion{Fe}{xii} {(193\,\AA, $\sim1.5\times 10^6$\,K)} line making use of Chianti v7 \citep{landi2013}, and in particular the coronal abundances of \cite{schmelz2012} -- see Figure \ref{fig:braid}(b).
%
%
Under the assumption of ionization equilibrium, at each grid point the emission to be expected from the \ion{Fe}{xii} line at 193\,{\AA} is calculated based on the atomic data and procedures contained in Chianti. The spectral profile at each grid point is then assumed to be a Gaussian with a width being the thermal width as defined in Eq.\,(\ref{E:th.width}) with the temperature $T$ given as found in the MHD model at that grid point and the mass $m$ of the Fe atom. The profile at each grid point is shifted by a wavelength equivalent to the Doppler shift caused by the velocity along the line-of-sight at that gridpoint. Finally, the line profiles are integrated along the line-of-sight.

As the loop evolves and magnetic energy is converted to kinetic and thermal energy, the emission in the coronal line of \ion{Fe}{xii} evolves, too.
As shown by the blue curve in Figure \ref{fig:lw_em}(a), the loop brightens and then dims in the \ion{Fe}{xii} emission. Looking at the black curves in the same plot, we see that this is due to a heating of loop strands to $\sim1.5\times10^6$ K, followed by a subsequent cooling. Note that the radiative cooling time (a few hundred seconds at these parameters) is longer than the timescale of the observed rapid drop in temperature \cite[and in addition the full radiative losses are not applied -- see][]{pontin2017}. Comparing the peak temperature (solid line) and average temperature (dashed), we see that the cooling is primarily due to a redistribution of the thermal energy. This is achieved by (i) thermal conduction along the loop, and (ii) transport of heat perpendicular to the loop axis by the continual reconnection of field lines \citep[every field line is on average reconnected multiple times during the relaxation, see][]{pontin2011a}.

\section{Line profiles {in the forward model}\label{sec:line}}

{%
In the following we will present spectral profiles resulting from our forward model.
We will investigate if the four key observational properties concerning non-thermal broadening are reproduced by our model.
As detailed in the introduction, these are that the non-thermal broadening
(i) is typically about 15--30\,\kms,
(ii) is independent of spatial resolution, and
(iii) shows a correlation with intensity.
Finally, (iv) the line profiles show enhanced wings.
As discussed in the introduction, these properties are commonly found in all major emission lines from the transition region and corona in a number of solar features.
For the purpose of this study, we will concentrate on the properties found in coronal emission, more specifically on the \ion{Fe}{xii} that forms at around 1.5\,MK in equilibrium.
For such a coronal line in active region loops (as we model here) the non-thermal broadening is at the lower end of the above mentioned range, about 15--20 \,{\kms} \cite[e.g.][]{2016ApJ...827...99T}. 
This choice of \ion{Fe}{xii} is motivated because many of the observations we refer to in the introduction have been performed with lines from this ion, and because it is a typical coronal line widely used in coronal studies --- in spectroscopy as well as imaging.
Because our model does not include a transition region from the chromosphere to the corona, any conclusions for lines forming at temperatures below 1\,MK have to wait for a more advanced model.
}%

\subsection{Line width: loop-averaged behaviour\label{sec:width.avg}}
\begin{figure*}
\centering
\includegraphics[width=0.99\textwidth]{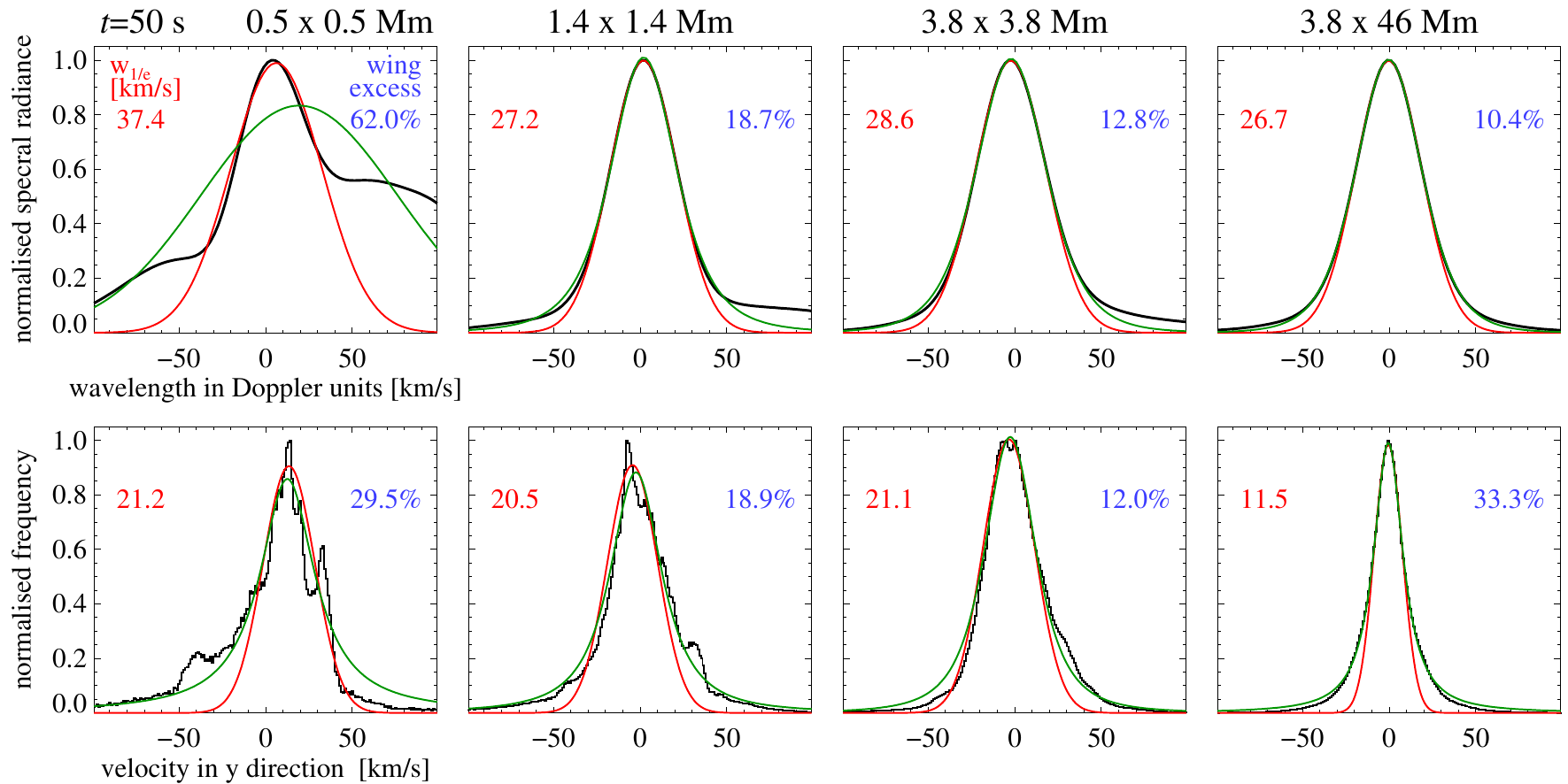}
\caption{Synthesised line profiles and velocity distributions at time $t=50$\,s.
Top row: Profiles of the \ion{Fe}{xii} line at 193\,{\AA} with the peak intensity normalized to unity plotted as a function of Doppler velocity (black).
These are for field-of-views of different sizes as indicated above the panels and illustrated in Fig.\,\ref{fig:braid}(b).
%
Bottom: Histograms of LOS velocities ($v_y$) at voxels within the corresponding viewing window (black).
The red lines show single Gaussian fits to the respective profiles. The numbers in red in the top left of each panel indicate the width of each Gaussian fit (Gaussian width equal to half width at $1/e$ of line peak).
The percentages in blue in the top right show the emission of the actual profile (in black) in excess of the single Gaussian fit (in red).
The green profiles show a fit to the line profile or velocity distribution by a kappa function. 
}
\label{fig:lineprofiles}
\end{figure*}
{We first investigate the average properties of the synthetic line profiles of \ion{Fe}{xii} synthesised as described in Sect.\,\ref{sec:simulations}.}
We focus primarily on profiles obtained for line-of-sight (LOS) perpendicular to the loop axis, along which the strongest Doppler velocities are found. We find that the particular LOS perpendicular to the loop axis chosen does not materially affect the conclusions, and so we stick here to an LOS parallel to the $y$-axis.

First we take a viewing window that covers a substantial potion of the simulated loop, and plot the width of the spectral  line as a function of time -- see Figure \ref{fig:lw_em}(b). (Specifically, at each time we fit a Gaussian to the line profile, and extract the $1/e$-width of the Gaussian.) We clearly observe a rapid increase in the line width above the thermal width over the first $\sim20$ s, after which the line width slowly decreases again towards the thermal value, in line with the decaying nature of the turbulence during the relaxation (as evidenced by the magnetic energy curve in the Figure). 
The thermal speed for \ion{Fe}{xii} in the corona is approximately $21$ km/s (assuming a line formation temperature of $T\approx 1.5\times10^6$ K). 
The line width drops to roughly this value after the magnetic field is relaxed at around 200\,s to 300\,s and hence the turbulent motions become negligible. At that time the non-resolved motions within the loop have ceased, meaning the non-thermal broadening dropped to zero, and the line width dropped to the thermal width.
We also find an oscillatory behaviour in the line width which is on the timescale of the Alfv{\'e}n loop travel time of 50 s and essentially reflects waves bouncing back and forth through the loop.

With the significant temporal evolution of the line width, the question arises which width should be representative, i.e.\ which width we would expect in actual observations based on this model.
The intensity in the loop reaches a peak value around $t=100$\,s to 140\,s (larger than 90\% of peak value; Fig.\,\ref{fig:lw_em}a).
Consequently, the loop element modeled here would be best visible (and probably brighter than its surroundings) during that time frame.
Thus the width at these times should be expected in observations.
Based on Fig.\,\ref{fig:lw_em}b we find $1/e$ widths of the \ion{Fe}{xii} line ranging from about $25$\,km/s to 30\,km/s.
Subtracting the thermal width of \ion{Fe}{xii} (21.2\,km/s), we find that these values correspond to non-thermal broadenings in the range of 14\,km/s to 21\,km/s (see Eq.~\ref{E:non-thermal}).
This coincides very well with the range of non-thermal broadening found {in observations for the very same line \cite[e.g.][]{2016ApJ...827...99T}.
Thus we can explain the key observational finding (i) as detailed in the introduction and the beginning of Sect.\,\ref{sec:line} --- at least} around the line formation temperature of \ion{Fe}{xii}, which is 1.5\,MK.

Due to the dominance of the magnetic field component along the loop, the r.m.s. value of $v_z$ is substantially smaller than $v_x$ and $v_y$ (see Figure \ref{fig:jvcuts}). As a result, the line broadening for a LOS parallel to the loop axis is substantially smaller. This is demonstrated by the dashed line in Figure \ref{fig:lw_em}, which shows the line width for a portion of the loop for $z\in[10,20]$ Mm (chosen to simulate the effect of looking down the leg of a curved loop). {In the Figure we see that the line width is in this case always less than or equal to the thermal width at line-formation temperature. This is because much of the plasma in the loop is well below the line-formation temperature, at $T<1$ MK, at which the thermal width is $w_{th}<17.2$ km/s.} It is worth noting that the average value of ${\bf B}_{xy}$ (relative to $B_z$) is expected to scale with $R_m$ \citep{longcope1994,ng2012}, so that for coronal parameters (for which the average of ${\bf B}_{xy}$ is expected to decrease more slowly during the turbulent decay) we might expect less of a discrepancy between the degree of the broadening between the perpendicular and parallel directions. {Furthermore, flows along the loop are also suppressed by the closed boundaries at the line-tied footpoints (at which the mass flux is zero), which might contribute to our under-estimation of the non-thermal broadening parallel to the loop axis.}

\subsection{Line width: dependence on resolution and field of view}
To probe the behaviour in more detail we select fields of view (FOV) of different sizes, and plot the line profile for these in Figure \ref{fig:lineprofiles}, at the time $t=50$\,s, i.e. at the first time that the loop-averaged loop emissivity exceeds 50\% of its temporal maximum. We choose three approximately square FOVs of increasing size, and a final FOV that covers the entire loop excluding the footpoints -- see the boxes marked in Figure \ref{fig:braid}(b). In the bottom row in the figure, histograms of the LOS velocity ($v_y$) in the voxels within the different FOVs are shown for comparison. 
We observe that the line width is largely unaffected by the size of the FOV (Figure \ref{fig:lineprofiles}). This is even more pronounced at later times (see plots in the appendix, Figs.\,\ref{fig:lineprofiles_t10} and \ref{fig:lineprofiles_t22}).

The reason for this independence of the line width on the {FOV size} can be found in the distribution of the velocities.
Because the turbulent motions are found on a large range of scales, down to the resolution limit of the simulation, when considering volumes with side length being a sizable fraction of the loop diameter, we can expect always roughly comparable widths of these distributions of turbulent velocities.
Looking at the velocity histograms, we find that the Doppler broadening is associated with a range of different LOS velocities along the LOS (note that the smallest FOV is only $0.5\times0.5$ Mm ($38\times6$ simulation pixels) compared with a depth along the LOS  of 4 Mm).
In general and very roughly, the width of the spectral line for each {FOV size}, $w_{\rm{line}}$, is given by the quadratic sum of the width of the velocity distribution $w_{\rm{velo}}$ and the thermal width of \ion{Fe}{xii} of about $w_{\rm{th}}=21$\,km/s, i.e. $w_{\rm{line}}^2\approx w_{\rm{velo}}^2 + w_{\rm{th}}^2$. Because the distribution of the turbulent velocities depends only weakly on the {FOV size} (when considering a sizable fraction of the loop diameter), also the width of the spectral line profiles would be insensitive to {FOV size}.

This is consistent with the key observational finding (ii) that the line broadening in the corona is independent of instrument resolution.
{Here as in the observations by \cite{2016ApJ...827...99T} this is the case for \ion{Fe}{xii}.}
The location of the centre of the FOV does not materially affect this result: although the details of the velocity histogram vary between different FOV locations over the loop, the overall broadening remains largely unchanged. Figures \ref{fig:lineprofiles_t10} and \ref{fig:lineprofiles_t22} in the Appendix show the gradual narrowing of the spectra and associated velocity pdfs at later time, consistent with the black curve in Figure \ref{fig:lw_em}(a).

\subsection{Non-Gaussian nature of line profiles -- enhanced wings}

%
Another striking feature of the line profiles in Figure \ref{fig:lineprofiles} is their systematically non-Gaussian nature.
To illustrate the non-Gaussian nature, we applied a single Gaussian fit to the line profiles. Because the lines appear to have excess emission in the line wings, for the fit we give more weight to the line core in order to get a good fit of the synthetic profiles near their center. In the plots, we note the width of the single Gaussian fit and the excess emission of the synthesized profile when compared to the single Gaussian fits (cf. top row of Fig.\,\ref{fig:lineprofiles}). The latter quantifies the non-Gaussian nature of the synthesized line profiles.

There is a clear enhancement in the wings of the distribution compared with the Gaussian fit.
Typically, the excess emission in the wings of the synthesized \ion{Fe}{xii} profiles (as compared to a single Gaussian) amounts to some 10\% to 20\%, even though for small regions this can be significantly higher (top row of Fig.\,\ref{fig:lineprofiles}). 
{This is qualitatively and quantitatively consistent with the key observational finding (iv), even though these symmetrically enhanced wings have only been reported in \ion{Fe}{xv} and \ion{Si}{iv}.
}

The non-Gaussian nature of the line profiles is a direct consequence of the LOS velocities, consistent with the turbulent nature of the evolution (as discussed in the following section).
The distributions of the velocities, when considered over a large enough region of more than 1$\times$1\,Mm$^2$, show also clearly enhanced wings (see bottom row of Fig.\,\ref{fig:lineprofiles}). Actually, these distributions can be fitted quite well by a kappa-distribution (over-plotted in green).

This enhancement in the wings is most pronounced at early times when the velocities are largest (in the first few 10s of seconds the evolution is dominated by a handful of reconnecting current sheets and associated flows, as the turbulent cascade is set up -- see e.g. \cite{pontin2011a} -- and so velocity histograms are dominated by a few sharp peaks for smaller FOVs).
After $\sim$50 seconds the current distribution has fragmented, and the velocity has become more turbulent -- the wing enhancements are quite pronounced at this time as shown in Figure \ref{fig:lineprofiles}. As time progresses, the additional power in the wings gradually reduces, and towards the end of the relaxation ($t>200$ s) -- when the turbulent decay is largely complete {(see Section \ref{sec:simulations})} -- the line profiles are indistinguishable from a fitted Gaussian (see plots in the appendix, Figs.\,\ref{fig:lineprofiles_t10} and \ref{fig:lineprofiles_t22}).
Essentially, as time goes by and the magnetic field relaxes, the line profiles reach a (single) Gaussian shape with a width comparable to the thermal width.

\begin{figure*}
\centering
\includegraphics[width=0.99\textwidth]{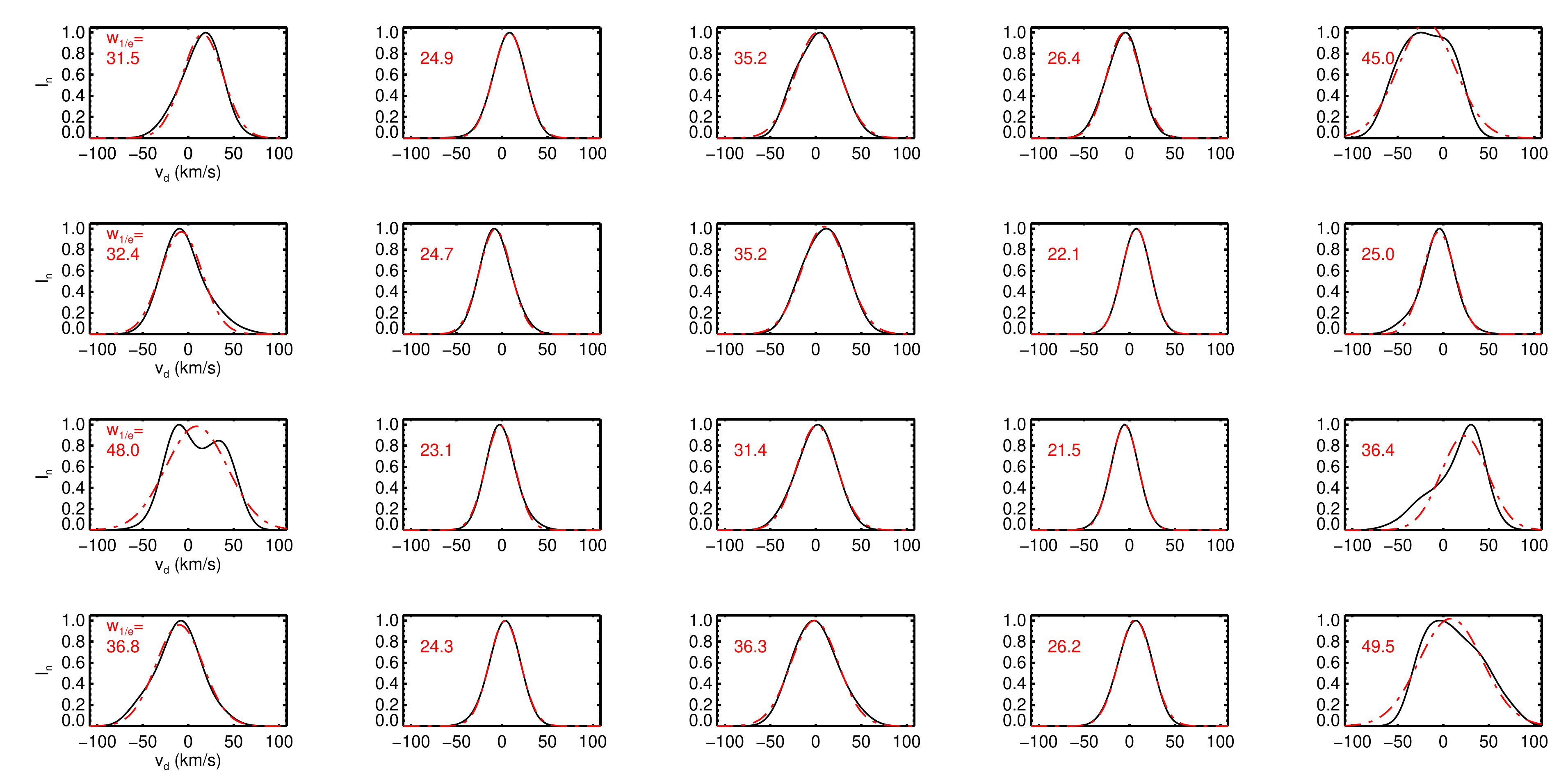}
\caption{Line profiles {of \ion{Fe}{xii}} for various times and windows of the same size as the smallest one in Figure \ref{fig:lineprofiles} (0.5Mm$\times$0.5Mm). Rows are times: $t=40$, $t=60$, $t=72$, $t=92$. Columns are position across the loop (all at same height): centre is in the centre and adjacent to it are windows adjacent on either side. The far-left and far-right plots are for windows located immediately inside the largest viewing window of Figures \ref{fig:braid}(b) and \ref{fig:jvcuts}.}
\label{fig:lineprofiles_small}
\end{figure*}
For the smallest viewing window selected, of size 0.5 Mm$^2$, the synthesised line profile is found to vary substantially between different locations in the loop and times, as shown in Figure \ref{fig:lineprofiles_small}. Comparing with the length scales of the velocities in Figure \ref{fig:jvcuts}, we see that individual coherent flows are  present at these scales. It is interesting to note that the broadest and most non-Gaussian profiles are to be found close to the edges of the loop (i.e.~away from the axis). This may be due to rapid twisting/untwisting of bundles of flux in these regions, though whether it is a generic feature would require further study of, e.g., different patterns of field braiding within the loop. Interestingly, at this size FOV we can start to pick out highly-non-Gaussian profiles including some with double peaks.


\subsection{Relation between line width and intensity\label{sec:corr.I.w}}

\begin{figure}
\centering
\includegraphics[width=0.49\textwidth]{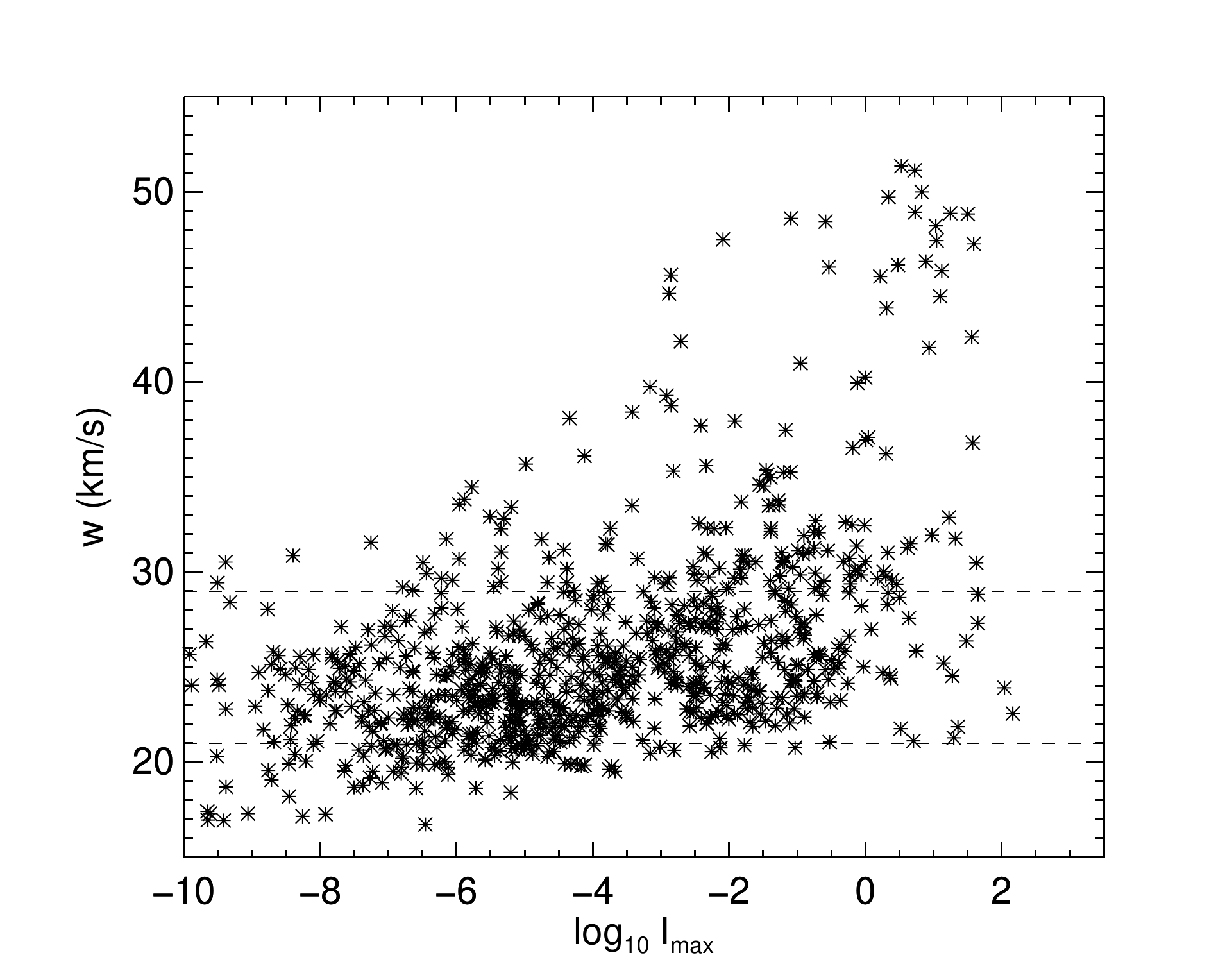}
\caption{Correlation between intensity and width {of \ion{Fe}{xii} synthesized from} the model. Scatter plot of the peak intensity versus the $1/e$-width of a fitted Gaussian, for viewing windows of size $0.5\,{\rm Mm}^2$ at $t=124$\,s. The horizontal lines at $w=21$ \kms and $w=29$ \kms correspond to non-thermal broadening of 0 and 20 \kms, respectively, based on a thermal width of 21 \kms ($T=1.5\times 10^6$ K).
See Sect.\,\ref{sec:corr.I.w}.}
\label{fig:int_vs_width}
\end{figure}

Finally, we examine the relation between the line width and the peak intensity for FOVs covering our entire loop. This is shown in Figure \ref{fig:int_vs_width} at the representative time of $t=124$\,s, when the loop is brightest (see Figure \ref{fig:lw_em}). We see, consistent with the observations, a clear correlation between the two.

Spectroscopic observations in the \ion{Fe}{xii} lines at 195\,\AA\ and 1349\,\AA\ show non-thermal motions in an active region loop of up to 40\,km/s, with the bulk part being in the range from zero to 20\,km/s non-thermal broadening \cite[][their Fig.\,10]{2016ApJ...827...99T}. This latter range corresponds to line widths (at $1/e$) from about 21\,km/s to 29\,km/s. We indicate this range in Figure \ref{fig:int_vs_width} by horizontal lines and find that indeed also in our model the bulk part of the data points shows widths {of \ion{Fe}{xii}} in this range. (Only at very small intensities we see line widths below the thermal width of \ion{Fe}{xii}, i.e. 21\,km/s, because these originate from cool plasma, below 1\,MK that would not show up in real observations). 
Equally important to matching the actual values of the line width, our model shows the trend that the line width increases with line intensity.

We can speculate that this correlation in our model is due to the fact that the strongest flows in the loop are associated with recently reconnected bundles of flux, and that these tend to have the highest temperatures due to Ohmic heating within the current layers \citep[see comparison of emission and current distribution in][]{pontin2017}.
This correlation is observed throughout the turbulent phase of the evolution, becoming much less pronounced at later times as the number of `pixels' with substantial non-thermal widths decreases.

\section{Discussion}\label{sec:discussion}

The results described above demonstrate that the non-thermal broadening in coronal lines is consistent with Parker's picture of coronal heating by magnetic field line braiding. 
%
%
To extract a ``typical" degree of broadening from the simulations, 
we consider the time period within which the loop would be most readily observable, specifically when the overall intensity of the loop is greater than $\sim90\%$ of its temporal maximum.
During this time the $1/e$-width of the \ion{Fe}{xii} emission averaged over the loop varies between 25 and 30\kms, corresponding to a non-thermal broadening between 14 and 21\kms, which is consistent with the observations as discussed in Section \ref{sec:intro}.
It is notable that the broadening is essentially independent of resolution and of the FOV location for FOV larger than 1.4 Mm$^2$ (0.5\arcsec$\times$0.5\arcsec), again consistent with observations {(see Section \ref{sec:intro})}. We speculate that this number maybe be dependent on the magnetic Reynolds number (though note that it is substantially larger than the typical grid spacing or current sheet thickness).



Another unexplained feature of observations {-- described in Section \ref{sec:intro} --} that is reproduced in our model is the non-Gaussian nature of the line profiles. Our study suggests that this is associated with the underlying nature of the velocity fluctuations.
In hydrodynamic turbulence, the non-Gaussian nature of velocity fluctuations has been demonstrated in experiments \citep{anselmet1984}. This is interpreted as a signature of ``intermittency" in the dynamics: a preferential occurrence of large events that violates the self-similarity across scales assumed in, for example, the original turbulence model of \cite{kolmogorov1941}. 
Such enhanced, non-Gaussian wings are also observed in probability distribution functions of various quantities in 2D and 3D MHD turbulence simulations \citep{biskamp2003pp}. Indeed, \cite{servidio2011} argued that the large events or coherent structures at dissipation scales are correlated with sites of current layers and magnetic reconnection. The non-Gaussian distribution of magnetic field and velocity fluctuations has also been observed in solar wind turbulence \citep{sorrisovalvo1999,bruno2013}.



It is worth mentioning here a number of restrictions of the simulations on which these results are based, whose relevance could be addressed in future work. First, there is no boundary driving in the simulations: this is consistent with the coronal evolution so long at the energy release time of the relaxing braid is short compared to the energy injection timescale. However, this is mostly likely not true for all coronal loops. For instance, the driving and energy injection could be more intermittent and bursty in coronal loops subjected to reconnection at their footpoints as found in recent observations \citep[e.g.][]{2017ApJS..229....4C,2018A&A...615L...9C}. Second, the dependence of the turbulent relaxation on magnetic Reynolds number, and effect on synthesised observables remains to be fully explored \citep[though see][]{pontin2011a}. Third, here the lower and upper boundaries of our domain are closed and line-tied, excluding the complex interaction with the lower layers of the atmosphere. Loop curvature (and associated expansion) and gravitational stratification are also ignored. Each of these restrictions is necessary for reasons of computational expense in order to sufficiently resolve the turbulent relaxation in the loop. However, their effect should be explored in future work.

\section{Conclusions}\label{sec:conc}
We have demonstrated that four key observational signatures of coronal loops are consistent with the braiding model for coronal heating first proposed by \cite{parker1988}. These are obtained by synthesising spectra based on MHD simulations of turbulent relaxation in a coronal loop, that is generated self-consistently by an observationally-motivated level of field line braiding.

First, non-thermal line widths from braiding-induced turbulence are the right order-of-magnitude to explain observed non-thermal line broadening.
Second, the fact that this broadening comes from line-of-sight integration (essentially independent of the field of view) may explain why broadening is observed to be independent of instrument resolution. Indeed, we find in the present simulations that the broadening is essentially independent of resolution or the particular location of the FOV for FOV sizes greater than 1.4 Mm$^2$. This number may well be influenced by the magnetic Reynolds number. 

Third, our simulations exhibit a {positive} correlation {between the} line intensity and the non-thermal broadening. We speculate that this correlation is due to the fact that the strongest flows in the loop are associated with recently reconnected bundles of flux, and that these tend to have the highest temperatures due to heating within the current layers.

Finally, the braiding-induced turbulence mechanism also provides a self-consistent explanation for wing enhancements on emission spectra. These are related to the non-Gaussian nature of the underlying velocity fluctuations that are responsible for the line broadening. This is associated with intermittency in the turbulence, consistent with previous observations and simulations of turbulence.

\appendix

\section{Additional plots}
Figures show the same plots as in the main body of the article, but at different times to demonstrate the temporal evolution. At later times as the turbulence dies out the spectra gradually return to a thermal width, and the excess wing broadening is substantially decreased.

\begin{figure*}
\centering
\includegraphics[width=0.99\textwidth]{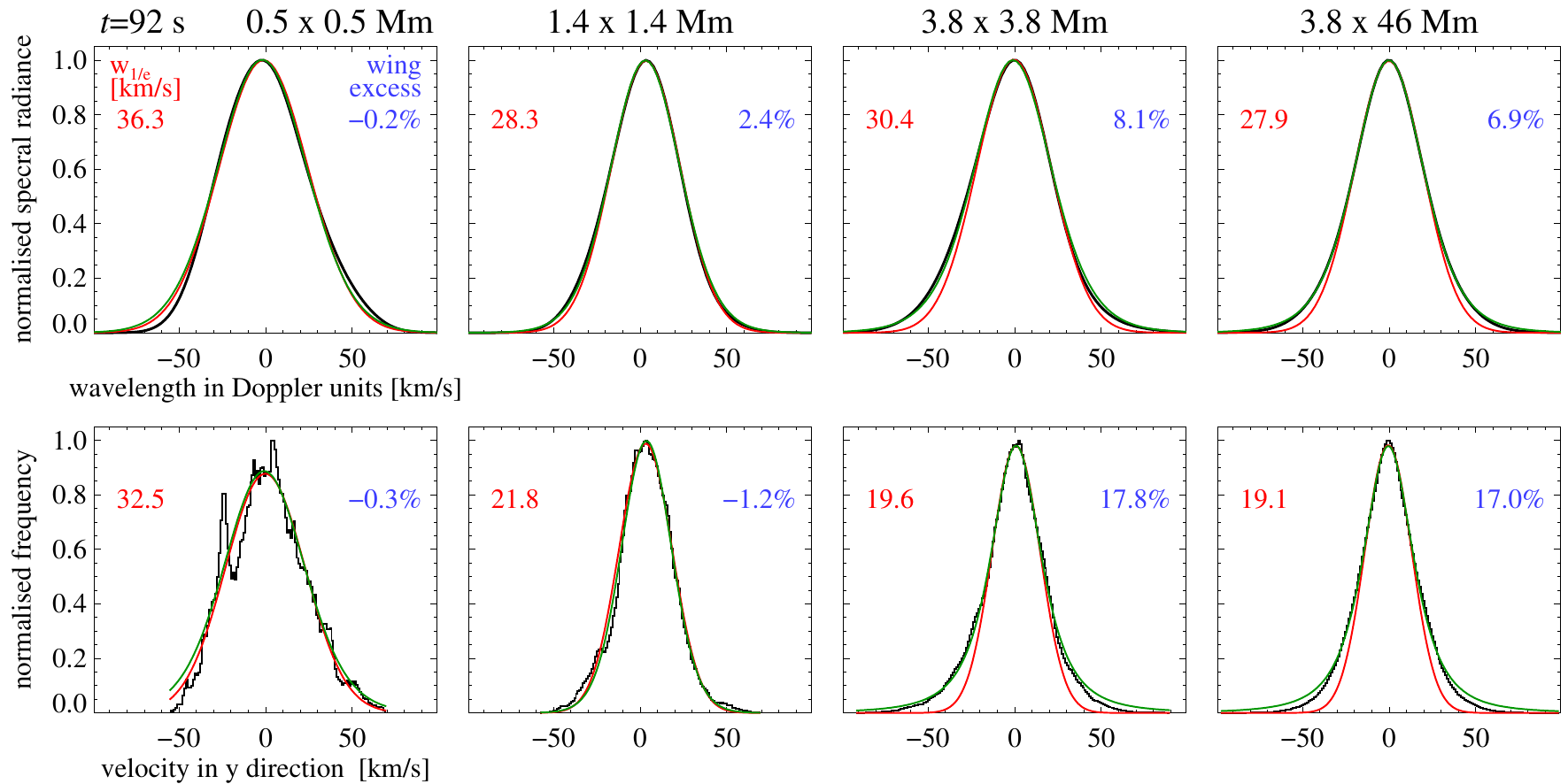}
\caption{Same as Figure \ref{fig:lineprofiles}, but at $t=92$ s.}
\label{fig:lineprofiles_t10}
\end{figure*}
\begin{figure*}
\centering
\includegraphics[width=0.99\textwidth]{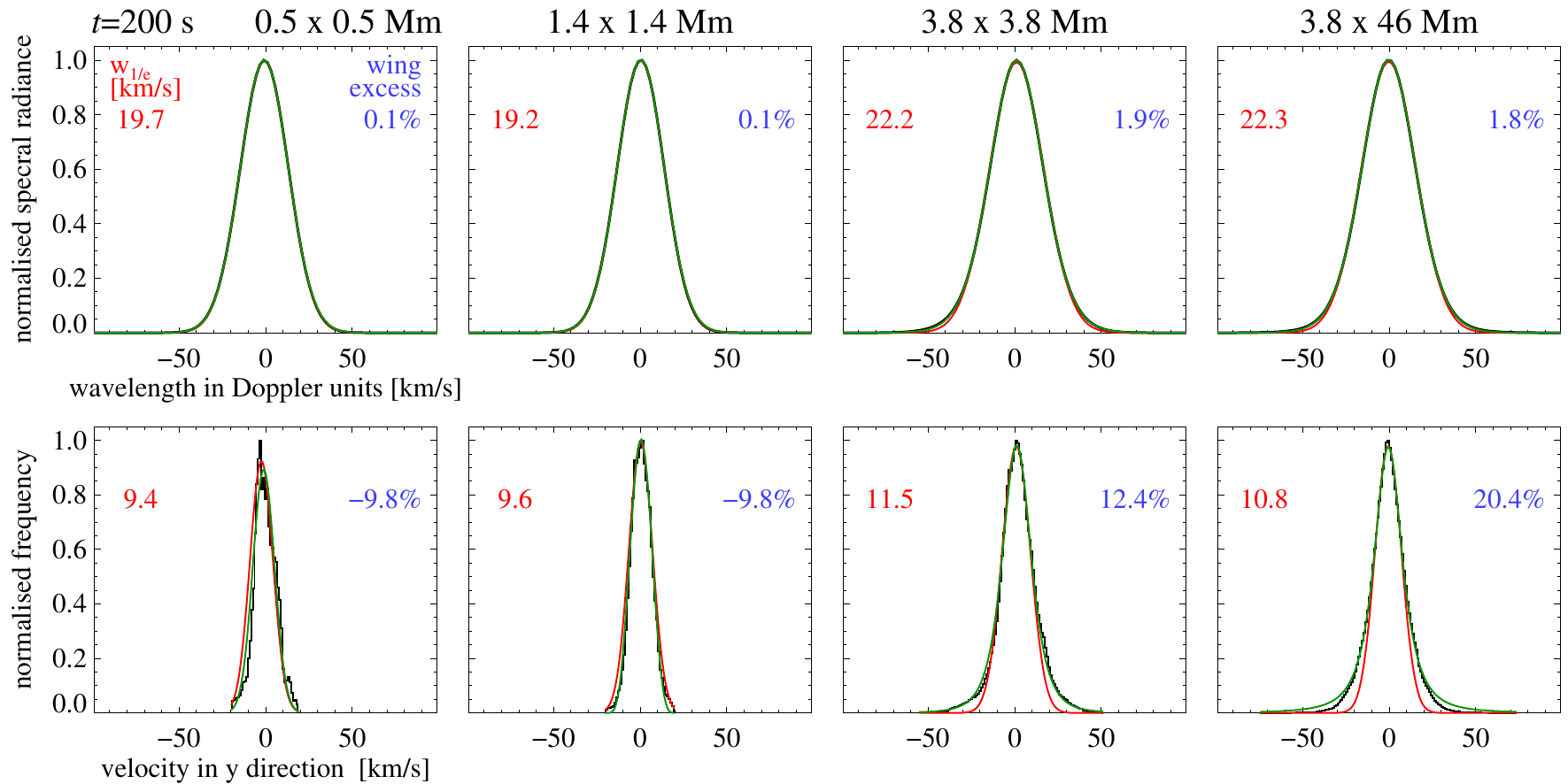}
\caption{Same as Figure \ref{fig:lineprofiles}, but at $t=200$ s.}
\label{fig:lineprofiles_t22}
\end{figure*}


\begin{acknowledgements}
DP gratefully acknowledges funding from the UK's STFC under grant ST/N000714. Simulations were run on the Cambridge Service for Data Driven Discovery, supported by an STFC DiRAC allocation to the UKMHD computing consortium. The authors acknowledge helpful discussions with Yi-Min Huang.
\end{acknowledgements}

%

%

\end{document}